\documentstyle [12pt]{ARTICLE}
\begin{document}
\def\DAF{DA$\Phi$NE\ }
\def\be{\begin{equation}}
\def\ee{\end{equation}}
\def\sd{\strut\displaystyle}
\begin{flushright}
UAB-FT-335/94\\
UG-FT-41/94\\
\end{flushright}
\vskip 1 cm
\begin{center}
{\bf EFFECTIVE CHIRAL LAGRANGIANS WITH AN SU(3)-BROKEN}
\vskip .4 cm
{\bf VECTOR-MESON SECTOR}
\vskip 1.5 cm
 A. Bramon \footnote{16414::BRAMON}\\
\vskip .1 cm
Grup de F\'\i sica Te\`orica, Universitat Aut\`onoma de Barcelona,\\
 08193 Bellaterra (Barcelona), Spain
\vskip .5 cm
A. Grau \footnote{16487::IGRAU}\\
\vskip .1 cm
Departamento de F\'\i sica Te\'orica y del Cosmos,\\
Universidad de Granada, 18071 Granada, Spain
\vskip .5 cm
and
\vskip .5 cm
G. Pancheri \footnote{PANCHERI $@$ IRMLNF}\\
\vskip .1 cm
INFN, Laboratori Nazionali di Frascati,\\
P.O.Box 13, 00044 Frascati, Italy
\vskip  2 cm
{\bf Abstract}
\end{center}
The accuracy of effective chiral lagrangians including
vector-mesons as (hidden symmetry) gauge fields is shown to be improved
by taking into account SU(3)-breaking in the vector-meson sector. The
masses, $M_V$, and couplings to the photon field and pseudoscalar
pairs, $f_{V\gamma}$ and $g_{VPP}$, are all consistently described in terms
of two parameters, which also fit  the values for the pseudoscalar
decay-constants, $f_P$, and charge radii, $<r^2_P>$. The description of
the latter is further improved when working in the context of chiral
perturbation theory.

\pagestyle{plane}
\setcounter{page}{1}
\newpage
\par

The question of including spin-1 mesons in effective chiral
lagrangians has been frequently discussed in the past
and has enjoyed considerable renewed interest during the
last few years. This is partly due to  the various proposals for
new low-energy,
high-luminosity $e^+e^-$ colliders which have been put forward and
approved, as in the  case
of  the \DAF $\phi$-factory, presently under construction
in Frascati.
\DAF
is expected to provide very  accurate and exhaustive experimental
information on $\phi$-decays, as well as on a number of low-energy
parameters, both in the kaon system and in other pseudoscalar and vector
 meson decays. This recent revival, spurred by the expectation for new
experimental data, has had
its counterpart on the theoretical front. Indeed, traditional ideas
concerned with Vector Meson Dominance, somehow associating spin-1 mesons
with gauge bosons of local
symmetries, have been further investigated and theoretically developed.
The so-called ``massive Yang-Mills
approach'' and ``hidden symmetry scheme'' recently reviewed
by Meissner \cite{Meissner} and Bando \cite{Bando} are two excellent
examples. The parallel developement of Chiral
Perturbation Theory (ChPT) \cite{GL} and its successful description of
the low-energy interactions of the pseudoscalar-meson octet --
a success considerably enlarged when spin-1 mesons are appropriately
taken into account \cite{dR} -- also points in the same direction.
Our purpose in the present note is to go
one step further by refining
the predictions of the above schemes through the introduction of a
well-known effect in hadron physics, namely, SU(3)-breaking
in the vector-meson sector,
a refinement that forthcoming data will certainly require.
\par
Conventional ChPT accounts for the electro-weak and strong interactions of
the pseudoscalar meson octet, $P$, in a perturbative series
expansion in terms
of their masses or four-momenta \cite{GL}.
At lowest order in this expansion, the
chiral lagrangian starts with the term
\be
\label{L2}
{\cal{L}}^{(2)}  =\sd {f^2\over 8} Tr({\cal{D}}_{\mu}\Sigma {\cal{D}}^{\mu}
\Sigma^\dagger)
\ee
where $\Sigma = \xi\xi = exp (2iP/f)$ and $f$ is the pion decay
constant $f=132\ MeV=f_{\pi}\ =f_K$, at this lowest-order level.
Electromagnetic interactions are contained in the covariant derivative
\be
{\cal{D}}_\mu \Sigma  = \partial_{\mu}\Sigma
+ieA_{\mu}[Q,\Sigma]
\ee
where $A_{\mu}$ is the photon field and $Q$ the quark-charge matrix (the
extension to weak interactions is trivial).
The mass degeneracy is broken via the additional mass term
\be
\label{L2m}
{\cal{L}}^{(2)}_m=\sd {f^2\over 8} Tr(\chi\Sigma^\dagger + \Sigma \chi^\dagger)
\ee
with $\chi$ containing the quark-mass matrix
$M=diag(m_u,m_d,m_s)$ and transforming
as a $(3,3^*)+(3^*,3)$ representation of $SU(3)_L \times SU(3)_R$. At
this lowest order, ChPT essentially coincides with Current Algebra.
\par
The next order piece of the chiral expansion (fourth order)
contains one-loop corrections
with vertices from (\ref{L2}) to (\ref{L2m}) and a series of ten counterterms
required to cancel loop divergencies \cite{GL}. Some of them, e.g.,
\be
{\cal{L}}^{(4)}_9 =-ie L_9 F_{\mu\nu} Tr(Q\ {\cal{D}}^\mu \Sigma\ {\cal{D}}^\nu
\Sigma^\dagger
+ Q\ {\cal{D}}^\mu \Sigma^\dagger \ {\cal{D}}^\nu \Sigma)
\ee
are chirally SU(3)-symmetric, whereas others, e.g.,
\be
{\cal{L}}^{(4)}_5 =\ L_5 Tr \left ({\cal{D}}_{\mu}\Sigma {\cal{D}}^{\mu}
\Sigma^\dagger \left(
\chi\Sigma^\dagger + \Sigma \chi^\dagger\right) \right)
\ee
break the symmetry as ${\cal{L}}^{(2)}_m$ in eq.(\ref{L2m}). At this one-loop
level, one obtains
\cite{GL}
\be
\label{fk/fpi}
\sd {f_K \over f_{\pi}} = 1+ {5\over 4}\mu_\pi-{1 \over 2}\mu_K -{3\over
4}\mu_\eta+
{8\over f^2}(m_K^2-m_\pi^2)L_5(\mu)
\ee
where loop effects appear through the so-called chiral-logs, $\mu_P= \sd
{m_P^2\over 16\pi^2 f^2}\ ln {m_P^2\over \mu^2}$.  Similarly, the pseudoscalar
electromagnetic charge-radii are found to be \cite{GL}
\begin{eqnarray}
\label{Ch-radii}
<r^2_{\pi^+}>&=&\sd {-1\over 16\pi^2 f^2}\left( 3+2\ ln {m_\pi^2\over \mu^2}+
 ln {m_K^2\over \mu^2}\right)+ {24\over f^2} L_9(\mu) \nonumber \\
<r^2_{K^+}>&=&\sd {-1\over 16\pi^2 f^2}\left( 3+\ ln {m_\pi^2\over \mu^2}+2\
 ln {m_K^2\over \mu^2}\right)+ {24\over f^2} L_9(\mu) \nonumber \\
-<r^2_{K^0}>&=&<r^2_{\pi^+}>-<r^2_{K^+}>\ =\ {1\over 16\pi^2 f^2}\ ln
{m_K^2\over m^2_\pi}
\end{eqnarray}
where, again, the divergencies accompanying the chiral-logs are absorbed
in one of the counterterms (or low-energy constants), $L_9$,
appearing in
${\cal{L}}^{(4)}$.
\par
The large number of low-energy constants, which need to be determined
along the lines
just discussed, considerably reduces the predictive power of ChPT, unless
one adopts the so called resonance saturation hypothesis\cite{dR}.
According to this suggestion,
 the values of these
constants are related to the exchange of the known meson resonances, so
far ignored in
the original context of ChPT.
This is a very attractive possibility
which has been  successfully verified in many cases. In
most of them, particularly in processes involving the Wess-Zumino
anomalous action \cite{Bijnens}, vector-mesons turn out to play the
dominant role.
Their couplings to pseudoscalar plus photon states, as extracted from
the experiments, saturate an important part of the above counterterms.
Accurate effective lagrangians incorporating additional properties
of vector mesons and further details on their dynamics could therefore
be extremely useful not only as a self-contained effective theory
but also as an auxiliary lagrangian fixing the counterterms in the
ChPT context.
\par
We follow the ``hidden symmetry approach'' of Bando et al.\cite{Bando}
to write the most general lagrangian containing
pseudoscalar, vector and
(external) electroweak gauge fields with the smallest
number of derivatives. At this lowest-order, it is given
by the linear combination
${\cal L}_A + a {\cal L}_V$, $a$ being an arbitrary parameter, of the
two lagrangians
\begin{eqnarray}
\label{LALV}
{\cal L}_A =\sd {-f^2\over 8} Tr  \left(
 D_{\mu}\xi_{L}\cdot \xi_{L}^{\dagger}- D_{\mu}\xi_{R}
\cdot \xi_{R}^{\dagger} \right)^2 \nonumber \\
{\cal L}_V  =\sd {-f^2\over 8} Tr \left(
 D_{\mu}\xi_{L}\cdot \xi_{L}^{\dagger}+ D_{\mu}\xi_{R}
\cdot \xi_{R}^{\dagger} \right)^2
\end{eqnarray}
both possessing a global (chiral) $SU(3)_L \times SU(3)_R$ and a local
(hidden) $SU(3)_V$ symmetry.
The $SU(3)$-valued matrices $\xi_{L}$ and $\xi_R$ contain the
pseudoscalar fields, $P$, and the unphysical (or compensator)
scalar fields, $\sigma$, that will be absorbed to give a mass to the
vector mesons
\be
\xi_{L,R}= \sd exp \left( i\sigma/f \right)\cdot
exp  \left( \mp iP/f \right)
\ee
The full covariant derivative is
\be
\label{Covariant}
D_{\mu} \xi_{L(R)}=\left (\partial_{\mu}-igV_{\mu}\right) \xi_{L(R)} +ie
\xi_{L(R)}
A_{\mu}\cdot Q
\ee
where, as before,  only the photon field, $A_{\mu}$,
has been explicitly shown, and $P$ and $V$ stand for the
SU(3) octet and nonet matrices
\be
P=\pmatrix{ \sd {\pi^0\over \sqrt 2}+{\eta \over \sqrt 6 }& \pi^+& K^+ \cr
               \pi^- &\sd -{\pi^0\over \sqrt 2}+{\eta \over \sqrt 6 } & K^0 \cr
                   K^- & \bar K^0  & \sd -{2\eta \over \sqrt 6}\cr} \qquad
V=\pmatrix{\sd {\rho^0\over \sqrt 2}+{\omega \over \sqrt 2 }&\rho^+& K^{*+} \cr
            \rho^-&\sd -{\rho^0\over \sqrt 2}+{\omega \over \sqrt 2 }&K^{*0}
\cr
                   K^{*-} & \bar {K^{*0}}  & \sd \phi \cr}
\ee
Under  $SU(3)_L \times SU(3)_R \times SU(3)_V$, the transformation
properties of $\xi_{L,R} (x)$ and the $SU(3)_V$ gauge-fields
$V_{\mu} (x)$ are $\xi_{L,R} (x) \to h(x) \xi_{L,R} (x)
g_{L,R}^{\dagger}$ and $V_{\mu} (x) \to i h(x) \partial_{\mu} h^{\dagger}(x) \
+ \ h(x) V_{\mu} (x) h^{\dagger}(x)$, where $g_{L,R} \in SU(3)_{L,R}$
and $h(x)$ parametrizes local, hidden $SU(3)_V$-transformations.
The lagrangian ${\cal L}_A + a {\cal L}_V$ can be reduced to the
chiral lagrangian (\ref{L2}) for any value of the parameter $a$ \cite{Bando}.
In fact, working in the unitary gauge, $i.e.$ fixing the local $SU(3)_V$ gauge
by $\xi_L^{\dagger} = \xi_R =\xi= \sd exp \left (iP/f
\right ) $ to eliminate the unphysical scalar fields, and
substituting the
solution of the equation of motion for $V_{\mu}$, the
${\cal L}_V$ part vanishes
and ${\cal L}_A$ becomes identical to the non linear chiral lagrangian
(\ref{L2}).

 \par
The ``hidden symmetry'' lagrangian ${\cal{L}}_A + a {\cal L}_V$ (\ref{LALV})
can
be easily seen to contain, among other things, a vector meson mass
term, the pseudoscalar weak decay constants, the vector-photon
conversion
factor and the couplings of both vectors and photons to pseudoscalar
pairs. The latter can be eliminated fixing $a = 2$, thus incorporating
conventional vector-dominance in the electromagnetic form-factors
of pseudoscalars. With this lagrangian one can
therefore attempt to describe the following sets
of data (that we take from \cite{PDG} neglecting error bars when
negligible):
\par\noindent
the vector meson mass spectrum
\be
\label{data-M}
M^2_{\rho,\omega},\ M^2_{K^*}, \  M^2_{\phi} =
0.60,\ 0.80,\ 1.04\ GeV^2;
\ee
the weak decay constants
\be
\label{data-f}
f_{\pi}= 132\ MeV,\ \ f_K= 160\ MeV;
\ee
the $\rho-\gamma, \omega-\gamma$ and $\phi-\gamma$ conversion factors,
$f_{V\gamma}$. This fixes the gauge coupling $g$, once the respective quark
charges in $V\to e^+e^-$ have been taken into account, to
\be
\label{data-g1}
g=3.6\pm0.1,\  4.0\pm0.1,\  4.0\pm0.1;
\ee
and the $g_{\rho\pi\pi}=\sqrt 2\ g$, $g_{K^*K\pi}=\sqrt 3\ g/\sqrt 2$ and
$g_{\phi K K}=\sqrt 2\ g$ decay constants, which lead similarly to
\be
\label{data-g2}
g=4.3,\  4.5,\  4.6
\ee
with negligible errors. From \cite{Olson} we take the following
experimental values for the electromagnetic (e.m.) charge radii
\begin{eqnarray}
\label{data-radii}
<r^2_{\pi^+}>&=&0.44\pm 0.03\ fm^2 \nonumber \\
<r^2_{K^+}>&=&0.31\pm 0.05\ fm^2 \nonumber \\
-<r^2_{K^0}>&=&0.054\pm 0.026\ fm^2
\end{eqnarray}
and the combined result, free from most systematic errors (see Dally et
al. and Amendolia et al. in ref.\cite{Olson})
\be
<r^2_{\pi^+}>-<r^2_{K^+}>=
0.13\pm 0.04\ fm^2
\ee
\par
The most immediate possibility to account for the above sets of
data consists
in using the ``hidden symmetry lagrangian'' (\ref{LALV})
as a self contained effective
theory in the exact SU(3) limit. In this case, the SU(3)-breaking
effects shown by some of the above data remain unexplained, but two
successful relations can be obtained for the non-strange sector.
The lagrangian (\ref{LALV}) predicts $M^2_{\rho,\omega} = 2g^2f^2$ in the
$0.50 - 0.60 \ GeV^2$ range when the values (\ref{data-f}, \ref{data-g1},
\ref{data-g2}) for
$f$ and  $g$ are used. This agrees with the value obtained from
the direct measurement of the
$\rho, \omega$ mass (\ref{data-M}). Moreover, it also agrees with
that coming from the pion charge radius, leading
in our approach to $M^2_{\rho}=6/<r^2_{\pi}> = 0.51\ GeV^2$.
Alternatively, one can
regard the lagrangian (\ref{LALV}) in a ChPT context, include the chiral-logs
from the loop corrections and saturate the finite part of the required
counterterms with (\ref{LALV}). This maintains the successful relation
$M^2_{\rho,\omega}=2g^2f^2$ and gives $L_9=f^2/4M^2_{\rho}$, with
a vanishing value for the
SU(3)-breaking low-energy counterterm $L_5$. With
this
value for $L_9$, and evaluating the  chiral-log correction at the
conventional value $\mu=M_{\rho}$,  (\ref{Ch-radii})  predicts
 $< r^2_{\pi}>\ = 0.46
\ fm^2$ in complete agreement with experiment (\ref{data-radii}).
\par
The next step is to include SU(3)-breaking
effects. Since the pseudoscalar sector is known to break the symmetry as in
eq.(\ref{L2m}), $i.e.$, proportionally to
$ Tr (\xi_{R} M \xi_{L}^{\dagger}+ \xi_{L} M \xi_{R}^{\dagger}) $,
we incorporate SU(3) symmetry breaking, as already attempted
in ref.\cite{Bando},
via a similar hermitian  combination
 $(\xi_{L} \epsilon \xi_{R}^{\dagger}+ \xi_{R} \epsilon
\xi_{L}^{\dagger})$  in both  ${\cal L}_A$ and ${\cal L}_V$
terms, $i.e.$,
\be
\label{LA}
{\cal L}_A +\Delta {\cal L}_A =\sd {-f^2\over 8} Tr \left \{ \left(
 D_{\mu}\xi_{L}\cdot \xi_{L}^{\dagger}- D_{\mu}\xi_{R}
\cdot \xi_{R}^{\dagger} \right)^2 \left [ 1
 + \left( \xi_{L} \epsilon_A \xi_{R}^{\dagger}+ \xi_{R} \epsilon_A
\xi_{L}^{\dagger}
 \right )\right ] \right \}
\ee
and
\be
\label{LV}
{\cal L}_V +\Delta {\cal L}_V =\sd {-f^2\over 8}
 Tr \left \{ \left( D_{\mu}\xi_{L}\cdot
\xi_{L}^{\dagger}+ D_{\mu}\xi_{R}\cdot \xi_{R}^{\dagger}\right)^2 \left [ 1
 + \left( \xi_{L} \epsilon_V \xi_{R}^{\dagger}+ \xi_{R} \epsilon_V
 \xi_{L}^{\dagger} \right )\right ] \right \}
\ee
The matrix $\epsilon_{A(V)}$ is taken to be $\epsilon_{A(V)} =
diag (0,0,c_{A(V)})$, where $c_{A,V}$
are the SU(3) -breaking real parameters to be determined. Notice that
the SU(3)-breaking terms, $\Delta {\cal L}_{A,V}$, are hermitian,
unlike those in ref.\cite{Bando}.
Working in the unitary gauge $\xi_L^{\dagger} = \xi_R = \xi= \sd exp
\left (iP/f
\right ) $, ($\sigma=0$), and expanding in terms of the pseudoscalar
fields, one observes
that the kinetic terms in ${\cal L}_A$ have to be renormalized. This
is simply achieved by
rescaling the pseudoscalar fields \cite{Bando}

\begin{eqnarray}
\label{P-renorm}
 \sd  \sqrt {1+c_A}\ K &\to&  K         \nonumber \\
 \sd  \sqrt {1+2c_A/3}\ \eta &\to& \eta
\end{eqnarray}
where an $\eta$-$\eta'$ mixing angle of -$19.5^0$ has been used for the $\eta$
case.
\par
The physical content of this new, SU(3)-broken ``hidden symmetry''
lagrangian (\ref{LA}) and (\ref{LV}) can now be easily worked out.
{}From $a \left ( {\cal L}_V + \Delta {\cal L}_V \right )$, we
obtain the conventional SU(3)-splitting for the
vector meson masses ($a=2$).
\be
\label{masses}
M_\rho^2=M_\omega^2=2 g^2 f^2, \qquad M_{K^*}^2= M_\rho^2 (1+c_V),
\qquad
 M_\phi^2=M_\rho^2 (1+2 c_V)
\ee
For the V-$\gamma$ couplings,
the corresponding part of the lagrangian is explicitly given by:
\begin{eqnarray}
\label{LVG}
{\cal L}_{V\gamma}& =& -  egf^2 A_{\mu}\ Tr \left [ \left \{Q,V^{\mu}\right \}\
\left ( 1+2 \epsilon_V\right )\right ] \nonumber \\
&=& \sd - {e M^2_{\rho,\omega} \over \sqrt2 g}\ A_{\mu}
\left [\rho^{0\mu} + {\omega^{\mu}\over 3} - (1+2c_V)
{\sqrt 2\over 3} \phi^{\mu}\right ]
\end{eqnarray}
The new terms in the lagrangian (\ref{LA},\ref{LV}) also induce an SU(3)
symmetry
breaking in the $g_{VPP}$ coupling constants.
One obtains
\begin{eqnarray}
\label{g}
g_{\rho \pi \pi}&=&\sqrt 2\ g \nonumber \\
g_{\rho KK}&=& g_{\omega KK}\ =\ \sd {g\over \sqrt 2}\ {1 \over 1+c_A}
\nonumber \\
g_{\phi KK}&=&\sd  \sqrt 2\ g\ {1+2c_V \over 1+c_A} \nonumber \\
g_{K^{*} K \pi} &=&\sd
{\sqrt3 \over \sqrt2}\ g\ {(1+c_V) \over \sqrt {1+c_A}}
\end{eqnarray}
where the $c_A$-dependence comes from the symmetry breaking in the
${\cal L}_A+\Delta {\cal L}_A $
lagrangian due to the renormalization of the pseudoscalar fields
(see eq.(\ref{P-renorm})). This redefinition of the
pseudoscalar fields also implies  symmetry breaking in the
pseudoscalar meson decay constants, namely,
\begin{eqnarray}
\label{fP}
 \sd f_K &=& \sqrt {1+c_A} \ f_{\pi} \nonumber \\
 \sd f_{\eta} &=& \sqrt {1+2c_A/3}\ f_{\pi}
\end{eqnarray}
\par
One can now attempt a description of the whole set of data (\ref{data-M}-\ref
{data-radii}) solely in
terms of the SU(3)-broken lagrangian (\ref{LA}) and (\ref{LV}).
This fixes
the values of the two new free parameters to $c_V=0.30$ and $c_A=0.45$.
The fit is completely satisfactory for the four sets of data quoted in
eqs.(\ref{data-M}) to (\ref{data-g2}).
For the
pseudoscalar charge radii, one gets from
eqs.(\ref{LVG}, \ref{g}, \ref{fP})
\begin{eqnarray}
\label{r2P}
<r^2_{\pi^+}>&=&\sd {6 \over M_{\rho}^2}   \nonumber \\
<r^2_{K^+}>&=&\sd {<r^2_{\pi^+}> \over (1+c_A)}\ {1\over 3}\left( 2+(1+2c_V)
{M^2_{\rho,\omega}\over M^2_\phi}\right) \nonumber \\
- <r^2_{K^0}>&=&\sd {<r^2_{\pi^+}> \over (1+c_A)}\ {1\over 3}\left( 1-(1+2c_V)
{M^2_{\rho,\omega}\over M^2_\phi}\right)
\end{eqnarray}
and the above values of $c_V$ and $c_A$ imply
\be
<r^2_{\pi^+}>=0.39\ fm^2, \qquad <r^2_{K^+}>=0.26\ fm^2, \qquad - <r^2_{K^0}>
=0.01\ fm^2
\ee
somewhat below , by one or two $\sigma$'s, the experimental data
(\ref{data-radii}). As
previously discussed, an alternative, more sophisticated possibility is
to use our SU(3)-broken lagrangian (\ref{LA},\ref{LV}) in conjunction with
ChPT.
This can only modify the predictions for $f_P$ (\ref{fP}) and
$<r^2_P>$ (\ref{r2P}) related to
processes without vector-mesons in the external legs. The chiral-logs
in eq.(6), evaluated at the conventional value $\mu=M_{\rho}$, account
now for some 35\% of the observed difference between $f_K$ and
$f_{\pi}$, thus requiring a smaller contribution from the $L_5$
counterterm and, hence, a smaller value for $c_A$. Accordingly, the
best global fit is now achieved by the slightly modified values
\be
\label{cvca}
c_V=0.28 \qquad \qquad c_A= 0.36
\ee
which preserve the goodness of the preceding fit for $M_V$,
$f_{V\gamma}$, $g_{VPP}$ and $f_P$, while improving the agreement in
the $<r^2_P>$ sector. Indeed, adding the chiral loop contributions to the
charge radii (\ref{r2P}), one obtains
\begin{eqnarray}
\label{rkrpidif}
<r^2_{\pi^+}> &=& \sd {-1\over 16\pi^2 f^2}\left( 3+2\ ln {m_\pi^2\over \mu^2}+
 ln {m_K^2\over \mu^2}\right)+ {6 \over M_{\rho}^2} \nonumber \\
- <r^2_{K^0}> &=&\sd {1\over 16\pi^2 f^2}\ ln {m^2_K\over m^2_\pi} +
{2\over 1+c_A} \left [ {1\over M^2_\rho}-{(1+2 c_V) \over M^2_\phi} \right ]
\nonumber \\
<r^2_{\pi^+}>-<r^2_{K^+}>&=&\sd {1\over 16\pi^2 f^2}\ ln {m^2_K\over m^2_\pi} +
{6\over M^2_\rho} - {2\over 1+c_A} \left [ {2\over M^2_\rho}+ {(1+2 c_V) \over
M^2_\phi}
\right ] \nonumber \\
&
\end{eqnarray}
The chiral-logs enhance the
previous predictions for $<r^2_P>$ and we find
\be
<r^2_{\pi^+}>= 0.46 fm^2, \qquad <r^2_{K^+}>= 0.32 fm^2 , \qquad
- <r^2_{K^0}>=0.043 fm^2
\ee
in better agreement with the data (\ref{data-radii}).

Notice that the last two expressions in eq.(\ref{rkrpidif}), which
explicitly contain
SU(3) breaking effects, considerably improve
the one-loop ChPT results \cite{GL} quoted in the
last line of eq.(\ref{Ch-radii}), namely $- <r^2_{K^0}> =
<r^2_{\pi^+}>-<r^2_{K^+}>
=0.036\ fm^2$, which was significantly below the  data (17).
This
improvement stems from the fact that in saturating the ChPT
counterterms with our SU(3)-broken lagrangian one goes beyond the
fourth order counterterms in ChPT, such as the SU(3)-symmetric counterterm
$L_9$, and introduces corrections from its SU(3)-breaking analogues
belonging to sixth order counterterm(s) in ${\cal{L}}^{(6)}$. This is
explicitly seen in eqs.(\ref{rkrpidif}), which reduce to the conventional ChPT
result (\ref{Ch-radii}) only in the exact SU(3) limit  $c_A=c_V=0$ and
$M^2_{\rho,\omega}=M^2_{\phi}$. We summarize the above  in
Table I, where we indicate the results obtained for the charge radii in the
three models so far discussed, $i.e.$, Chiral Perturbation Theory (ChPT)
and SU(3) broken
``Hidden Symmetry'' scheme (HSS) with and without chiral loops.

\vskip 5 mm
\centerline{{\bf Table I} (all data are in $fm^2$)}
\[ \begin{array} {|c|c|c|c|c|}\hline
    &exp.\cite{Olson}  &  ChPT \cite{GL}&SU(3)broken& SU(3) broken  \\
    &                  &                  & HSS       &HSS + Ch\ loops  \\
\hline
<r^2_{\pi^+}> &  0.44 \ \pm \ 0.03  &  0.44     &  0.39&0.46   \\
<r^2_{K^+}>   &  0.31 \ \pm \ 0.05  &  0.40     &  0.26&0.32   \\
-<r^2_{K^0}>  &  0.054\ \pm \ 0.026 &  0.036    &  0.01&0.043  \\
<r^2_{\pi^+}>-<r^2_{K^+}>   &    0.13\ \pm\ 0.04&  0.036 & 0.13&0.14   \\
\hline
\end{array} \]

\par
One can easily extend the above results for the e.m. charge radii of
pseudoscalars to include their weak analogue in $K_{l3}$ decays. Sirlin's
theorem \cite{Alberto}, valid up to first order in SU(3)-breaking,
and requiring
\be
\label{SirlinTh}
<r^2_{K\pi}>-<r^2_{K^0}> = \sd {1\over 2} <r^2_{\pi^+}>+ {1\over 2} <r^2_{K^+}>
\ee
provides the clue. The data (\ref{data-radii}) and the
experimental value \cite{PDG} $<r^2_{K \pi}>=0.36 \ \pm\ 0.02\ fm^2$
are fully compatible with Sirlin's theorem (\ref{SirlinTh}).
On the
other hand, one immediately sees that
the predictions of our first-order SU(3)-breaking
lagrangian verify  the theorem, thus providing a check of
 our calculations
and their automatic extension to the $K\pi$ case. From the
expression
\begin{equation}
<r^2_{K\pi}>={{1+c_V}\over{\sqrt{1+c_A}}}{{6}\over{M^2_{K^*}}}+
\ chiral \ loop \ contributions
\end{equation}
where the contribution from chiral loops is the one to be
 found in ref. \cite{GL}, we obtain
the very acceptable value  $<r^2_{K\pi}>=0.33\ fm^2$. The above equation
 can be compared with the chiral perturbation
theory result \cite{GL}
\begin{equation}
<r^2_{K\pi}>=\ chiral \ loop \ contributions \ + {{24}\over{f^2}}L_9(\mu)
\end{equation}
which gives   $<r^2_{K\pi}>=0.38\ fm^2$.

An extension of the present treatment to processes related to the anomaly
or Wess-Zumino lagrangian is  also possible and will be discussed
elsewhere. Here we simply note that introducing our value $c_A=0.36$,
eq.(\ref{cvca}), in eq.(\ref{fP}) for $f_\eta$, leads to $\Gamma(\eta \to
\gamma \gamma)=0.53$ KeV quite in line with the measured decay
width \cite{PDG}, i.e. $\Gamma(\eta
\to \gamma \gamma)=0.46 \pm 0.05$ KeV.
\par
In summary, we have presented a model for low energy hadronic
interactions in which SU(3)-breaking effects in the vector-meson sector
are introduced in chiral
lagrangians  including
vector-mesons as (hidden symmetry) gauge fields. By comparing a set of low
energy data with the model predictions, we have shown that it is possible to
extract values for the SU(3) breaking parameters which are very reasonable
 and which can lead to predictions for the electromagnetic
 charge radii of the pseudoscalar mesons in good agreement with
experimental data.
Further applications, primarily to the anomalous
sector of the lagrangian, are being studied and will be presented in a
forthcoming paper.

\vskip 1cm
{\bf Acknowledgements}
\par
Thanks are due to M. Lavelle for a critical reading of the manuscript.
This work has been supported in part by the  Human Capital and
Mobility Programme, EEC Contract \# CHRX-CT920026, and by
CICYT, AEN 93-0520(A.B.)
and AEN 93-0615(A.G.).


\vskip 2 cm


\begin{thebibliography}{99}
\bibitem{Meissner}
U.-G. Meissner, Phys. Rep. $\underline{161}$ (1988) 213.
\bibitem{Bando}
M. Bando, T. Kugo and K. Yamawaki, Nucl. Phys. $\underline{B259}$ (1985) 493,
and
Phys. Rep. $\underline{164}$ (1988) 217.
\bibitem{GL}
J. Gasser and H. Leutwyler, Nucl. Phys. $\underline{B250}$ (1985) 465, 517,
539.
\bibitem{dR}
G. Ecker, J. Gasser, A. Pich and E. de Rafael, Nucl. Phys. $\underline{B321}$
(1989)
311.\\
J.F. Donoghue, C. Ramirez and G. Valencia, Phys. Rev. $\underline{D39}$ (1989)
1947.
\bibitem{Bijnens}
J. Bijnens, Int. J. Mod. Phys. $\underline{A8}$ (1993) 3045 and references
therein.
\bibitem{PDG}
{\it Particle Data Group}, Phys. Rev. $\underline{D45}$ (1992) 1.
\bibitem{Olson}
A. Quenzer et al., Phys. Lett. $\underline{B76}$ (1977) 512.\\
E.B. Dally et al., Phys. Rev. Lett. $\underline{39}$ (1977) 1176.\\
E.B. Dally et al., Phys. Rev. Lett. $\underline{48}$ (1982) 375.\\
S.R. Amendolia et al., Phys. Lett. $\underline{B178}$ (1986) 435.\\
W.R. Molzon et al., Phys. Rev. Lett. $\underline{41}$ (1978) 1231.\\
C. Erkal and M.G. Olsson, J. Phys. G: Nucl. Phys. $\underline{13}$ (1987) 1355.
\bibitem{Alberto}
A. Sirlin, Phys. Rev. Lett. $\underline{43}$ (1979) 904; Ann. Phys.(N.Y.)
$\underline{61}$ (1970) 294.

\end{thebibliography}
\end{document}